\documentclass{article}

\pdfoutput=1

\usepackage[ansinew]{inputenc}
\usepackage[cyr]{aeguill}
\usepackage{xspace}

\usepackage{rotating}
\usepackage{verbatim}
\usepackage[english,frenchb]{babel}
\usepackage[hyphens]{url}
\usepackage{graphicx}

\title{Note de lecture:\thanks{Pour Natalia} \\ 'Climat: 15 v\'erit\'es qui d\'erangent'}
\author{Alexis Merlaud \\
	Insititut d'A\'eronomie Spatiale de Belgique 
	}

\begin{document}

\maketitle

\begin{abstract}

Ceci est une analyse du livre 'Climat: 15 v\'erit\'es qui d\'erangent', publi\'e par un collectif sous la direction scientifique du Pr Itsvan Mark\'o et incluant Anne Debeil, Ludovic Delory, Samuel Furfari, Drieu Godefridi, Henri Masson, Lars Myren, et Alain Pr\'eat. Nous montrons qu'au sujet de la science du climat, le livre contient trop d'approximations ou d'erreurs pour \^etre pris au s\'erieux. Le livre est cependant int\'eressant pour d'autres aspects, en particulier d'un point de vue \'epist\'emologique. Nous insistons sur l'information principale du livre, qui n'est pas exprimée assez clairement par les auteurs: il y a une forte incompatibilité pratique entre la croyance en l'efficacité du marché libre et la théorie scientifique du changement climatique anthropique.

---

This is a critical review of the book 'Climat: 15 v\'erit\'es qui d\'erangent', under the scientific supervision of Pr Itsvan Mark\'o, and co-authored by Anne Debeil, Ludovic Delory, Samuel Furfari, Drieu Godefridi, Henri Masson, Lars Myren, and Alain Pr\'eat. We show that regarding climate science, the book contains too many approximations or mistakes to be considered seriously. Nevertheless, the book is interesting in several aspects, in particular from an epistemological perspective. We also argue that the main take-home message is not stated clearly enough by the authors: there is a strong practical incompatibility between economic \emph{laissez-faire} and anthropogenic climate change.

\end{abstract}

\section{Introduction}
\label{sec:intro}

\begin{quote}

Nous savons, par expérience, que ce livre sera abondamment critiqué par des personnes qui ne l’auront pas lu. 

Itsvan Markó, Anne Debeil, Ludovic Delory, Samuel Furfari, Drieu Godefridi, Henri Masson, Lars Myren, et Alain Préat, \emph{Climat: 15 vérités qui dérangent}, Texquis, p. 177, 2013\cite{marko_2013}

\end{quote}

Dans \emph{Les marchands de doute}\cite{oreskes_2012}, Naomi Oreskes et Erik Conway pr\'esentent certains des climatosceptiques am\'ericains comme des fondamentalistes du march\'e libre li\'es aux groupes de pression ultralib\'eraux. Des scientifiques nostalgiques de la guerre froide agiraient en moines-soldats au service de l'idéologie du laissez-faire avec une constance remarquable pour empêcher l'adoption de mesures de régulation dans tous les domaines environnementaux: tabagisme passif, pluies acides, trous d'ozone, et actuellement changement climatique. Les auteurs apportent de nombreux éléments pour étayer leur thèse et elle s'applique assez bien aux cas particuliers qu'ils décrivent, à l'instar de Fred Singer et du Heartland Institute. Il est cependant évident à ceux qui se sont intéressés aux débats sur le climat en Europe que tous les scientifiques qui contestent les conclusions du GIEC ne sont pas des fanatiques du marché libre. Dans \emph{La Recherche}\cite{recherche_2013}, H\'el\`ene Guillemot et Stefan Aykut distinguent trois d\'ebats sur le climat. Le premier d\'ebat est celui des \emph{marchands de doute} et ce livre est cité. Ces controverses sont vigoureuses aux États-Unis o\`{u} les marchands de doute ont contribué à l'absence de r\'eaction publique sérieuse, mais le contexte est différent en Europe. Le deuxième d\'ebat pr\'esent\'e par Guillemot et Aykut remonte \'a l'appel d'Heidelberg\cite{kaufman1992} et oppose ceux qui \og défendent la modernit\'e technoscientifique et accusent leurs adversaires d'\^{e}tre influenc\'es par une id\'eologie ``verte''\fg. Claude Allègre est cité en exemple. Le troisième d\'ebat est port\'e par \og des scientifiques de domaines ``frontaliers'' du climat \fg, qui critiquent en particulier le recours à la modélisation.

L'ouvrage \emph{Climat: 15 vérités qui dérangent}\cite{marko_2013}, écrit sous la direction du professeur Markó, évoque d'abord le livre de Claude Allègre\cite{allegre2010}. Comme Allègre, Markó et al. prétendent révolutionner la climatologie en substituant l'opinion publique à l'avis des experts. Les faits et la bibliographie sont traités avec la même légèreté par Markó et al. qu'ils l'\'étaient par Allègre. Comme pour ce dernier, il suffit de vérifier les références des affirmations les plus extravagantes pour comprendre leur supercherie\cite{huet2010}. A propos des graphiques, plusieurs d'entre eux proviennent de blogs, sur lesquels, rappelons-le, n'importe qui peut publier n'importe quoi.  Il est exact qu'aucune courbe de Markó et al. n'est redessinée dans le sens qui convient comme dans le livre d'Allègre, mais une légende est modifiée pour servir la thèse du livre (voir sect.~\ref{sec:keeling}), ce qui revient au même. Ce livre déçoit ou trompe le lecteur qui s'intéresse au climat, selon son niveau d'expertise et de crédulité. Mais tout est intéressant pourvu qu'on le regarde assez longtemps. 

Cette note de lecture s'adresse \`{a} un public scientifique au sens large, y compris et surtout aux non spécialistes du climat. Dans la section~\ref{sec:contreverites} de cette note, nous exposons 15 contreverit\'es de Markó et al. La liste n'est pas exhaustive mais ces exemples sont choisis pour leurs intérêts pédagogiques. Nous verrons que les auteurs, qui se présentent comme sceptiques, ont besoin de croire en fait à des propositions très peu étayées par la science. Nous nous intéressons dans la section~\ref{sec:epistem} à l'épistémologie particulière de l'ouvrage. Dans la section~\ref{sec:sources}, nous classons l'ouvrage de Markó et al. dans les trois débats mentionnés plus haut, en nous appuyant en particulier sur la bibliographie. Si les catégories de Guillemot et Aykut sont poreuses, il apparaît que l'ouvrage s'inscrit assez bien dans le travail des marchands de doute. La section~\ref{sec:externalites} développe donc l'antagonisme pratique entre th\'eorie de l'efficacité des march\'es libres et théorie du changement climatique anthropique et les problèmes environnementaux en général.

\section{Quinze contrev\'erit\'es qui arrangent \\Markó et al.}
\label{sec:contreverites}

\begin{verse}
 \begin{flushright}

Ces naufrageurs na\"{i}fs arm\'es de sarbacanes \\
Qui sacrifient ainsi au culte du cargo \\
En soufflant vers l'azur et les a\'eroplanes.\\ 
 \medskip
Serge Gainsbourg, Cargo culte, 1971

 \end{flushright}

\end{verse}

\subsection{Contrev\'erit\'e: la th\'eorie du changement \\climatique anthropique est bas\'ee sur des modèles}

\begin{quote}
La th\'eorie du changement climatique d\^{u}  \`{a} l'homme  se base sur des mod\`eles ou simulations num\'eriques [\ldots]	 \ [S]i l'algorithme est \'ecrit de fa\c{c}on à ce qu'une augmentation de CO$_2$ atmosph\'erique induise une \'el\'evation de la temp\'erature globale, alors la temp\'erature globale pr\'edite augmentera automatiquement lorsque la teneur en CO$_2$ de l'atmosph\`ere s’accro\^itra, quel que soit le degr\'e de sophistication du mod\`ele. Il est donc \'evident que ce r\'esultat ne prouve strictement rien. (Markó et al. 98,100)
\end{quote}

La th\'eorie du changement climatique d\^{u}  \`{a} l'homme remonte \`{a} Svante Arrhenius (1859-1927) \cite{arrhenius_1897,arrhenius_1908},  prix Nobel de Chimie en 1903.  Arrhenius s'inspirait des travaux ant\'erieurs de John Tyndall (1820-1893) \cite{tyndall_1827} et de Joseph Fourier (1768-1830) \cite{fourier_1827}. Arrhenius n'a jamais vu d'ordinateur. La th\'eorie est bas\'ee sur les observations reproductibles  du spectre infrarouge du CO$_2$  et du gradient de temp\'erature dans la basse atmosph\`{e}re, et sur des consid\'erations thermodynamiques de base.  Ceci n'est expliqu\'e nulle part dans le livre de Markó et al. C'est pourtant cette physique, et non les r\'esultats des mod\`{e}les, qui fait croire aux climatologues que le CO$_2$ a un effet sur la temp\'erature. Le lecteur intéressé par une description simple de la physique de l'effet de serre pourra consulter par exemple l'ouvrage de référence: \emph{Introduction to Atmospheric Chemistry}\cite{jacob_1999}.

\subsection{Contrev\'erit\'e: le GIEC suppose que toute \\l'augmen\-tation de CO$_2$ depuis la révolution \\ industrielle est d'origine fossile}

\begin{quote}La th\'eorie du GIEC est bas\'ee sur l'hypoth\`ese que l'augmentation
du CO$_2$ dans l'atmosphère depuis la p\'eriode pr\'eeindustrielle,
soit environ 20 \% dans les années quatre-vingt-dix, est enti\`erement
due aux combustibles fossiles. (Markó et al. 57)
\end{quote}

Voici un extrait de l'\emph{Assessment Report 4} (AR4) du GIEC\cite{IPCC2007} qui contredit cette affirmation de Markó et al. (c'est moi qui souligne):

\begin{quote}
Past emissions of \textbf{fossil fuels} and \textbf{cement production} have likely contributed
about \textbf{three-quarters} of the current RF [radiative forcing], with the remainder caused by \textbf{land use changes.}
\end{quote}

\subsection{Contrev\'erit\'e: la courbe en crosse a \'et\'e \'elimin\'ee \\de l'AR4}
\label{sec:mann}

\begin{quote}
Finalement, la fameuse courbe en crosse de hockey, qui passait
sous silence l'Optimum m\'edi\'eval (p\'eriode r\'ecente la plus
chaude, approximativement de 1160 à 1300 ans) et le Petit \^age
glaciaire (p\'eriode r\'ecente la plus froide entre 1500 et 1700 ans)
a \'et\'e \'elimin\'ee du rapport du GIEC 2007. (Markó et al. 76)
\end{quote}

\begin{sidewaysfigure}[ht]
\begin{center}
\includegraphics[width=16cm]{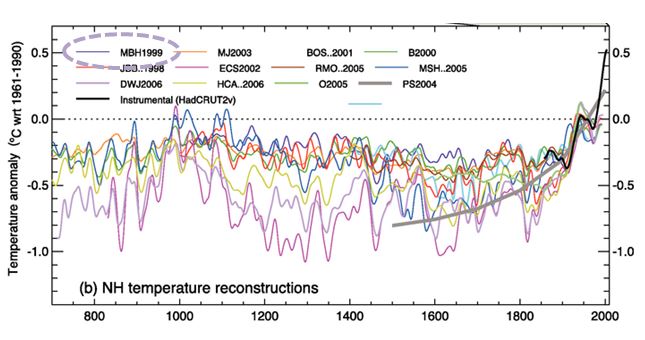}
\end{center}
\caption{Extrait de la figure 6.10.1 de l'AR4 montrant plusieurs reconstructions de temp\'erature dont MBH1999 (la courbe en crosse de Hockey), qui a \'et\'e supprimée de l'AR4 d'apr\'{e}s Markó et al. Les 11 autres reconstructions, bas\'ees sur diff\'erents proxies, soutiennent le r\'esultat initial de Mann et al.: le r\'echauffement au XXe si\`{e}cle est sans pr\'ecedent dans le dernier mill\'enaire.
}
\label{fig:mannAR4}
\end{sidewaysfigure}

La courbe en crosse de hockey a \'et\'e publi\'e par Mann et al. en 1999\cite{mann_1999} et figure dans le \emph{Third Assessment Report} (TAR) du GIEC\cite{IPCC2001} de 2001.  La m\'ethode statistique de Mann et al. a \'et\'e critiqu\'e \cite{mcintyre_2005}. Mann et al. ont publi\'e un corrigendum dans lequel ils reconnaissent une erreur minime\cite{mann_2004} et sans effet sur les r\'esultats. Dans l'AR4, publiée en 2007, cette courbe (MBH1999) est bien pr\'esente contrairement a ce qu'affirme Markó et al. Elle apparaît en effet dans la figure 6.10.1 de l'AR4\cite{IPCC2007}, dont un extrait est reproduit ici dans la figure~\ref{fig:mannAR4}. Cette figure est paradoxalement pr\'esente dans le livre de Markó et al., qui y reconnaissent m\^{e}me la courbe en crosse. Ce n'est pas la seule contradiction logique de Markó et al. (c.f. sections~\ref{sec:450} et ~\ref{sec:epistem}).   Les 11 autres reconstructions de la figure~\ref{fig:mannAR4}, bas\'ees sur diff\'erents proxies, soutiennent le r\'esultat initial de Mann et al.: le r\'echauffement au XX$^e$ si\`{e}cle est sans pr\'ecédent dans le dernier mill\'enaire. Voici un extrait de l'AR4:

\begin{quote}
The weight of current multi-proxy evidence, therefore, suggests greater 20th-century warmth, in comparison with temperature levels of the previous 400 years, than was shown in the TAR. On the evidence of the previous and four new reconstructions that reach back more than 1 kyr, it is likely that the 20th century was the warmest in at least the past 1.3 kyr.
\end{quote}

\subsection{\textbf{Contrev\'erit\'e: la divergence de Briffa n'a \\pas \'et\'e prise en consid\'eration par le GIEC.}}

\begin{quote}
\textbf{La forte divergence de la courbe de Briffa par rapport aux donn\'ees de la courbe
en crosse de hockey n'a jamais \'et\'e prise en consid\'eration dans les travaux du GIEC.} (Markó et al., 78, soulign\'e par les auteurs)
\end{quote}

La divergence \'evoqu\'ee par Markó et al. est celles des reconstructions de temp\'erature bas\'ees sur les cernes d'arbres par rapport aux temp\'eratures observ\'ees. Cette divergence \'etait d\'ej\`a connue avant le TAR\cite{jacoby_1997}. Contrairement \`a ce qu'affirment Markó et al., les reconstructions de Briffa et leurs limitations sont discut\'ees ouvertement dans l'AR4 (c'est moi qui souligne): 

\begin{quote}
Several analyses of ring width and ring density chronologies, with otherwise well-established sensitivity to temperature, have shown that they do not emulate the general warming trend evident in instrumental temperature records over recent decades, although they do track the warming that occurred during the early part of the 20th century and they continue to maintain a good correlation with observed temperatures over the full instrumental period at the interannual time scale (Briffa et al., 2004; D’Arrigo, 2006). This ‘divergence’ is apparently restricted to some northern, high latitude regions, but it is certainly not ubiquitous even there. In their large-scale reconstructions based on tree ring density data, \textbf{Briffa et al. (2001) specifically excluded the post-1960 data in their calibration against instrumental records, to avoid biasing the estimation of the earlier reconstructions (hence they are not shown in Figure 6.10)}, implicitly assuming that the ‘divergence’ was a uniquely recent phenomenon, as has also been argued by Cook et al. (2004a). (IPCC, AR4, 472,473)
\end{quote}


\subsection{Contrev\'erit\'e: le CO$_2$ n'augmente jamais \\ avant la temp\'erature}
\label{sec:keeling}

\begin{quote}
Les observations montrent que l'augmentation de la teneur atmosph\'erique en CO$_2$ est une cons\'equence de l'augmentation de la temp\'erature, et non l'inverse. \ldots \ Dans aucun des cas, on n'a pu constater que la hausse du CO$_2$ pr\'ec\'edait la hausse de la temp\'erature. (Markó et al., 95, 97)
\end{quote}

Deux graphiques viennent appuyer cette contrev\'erit\'e dans Markó et al. Le premier graphique (non reprdoduit ici) montre un d\'ephasage entre le CO$_2$ et la temp\'erature dans les carottes de glace de Vostok \`a l'\'echelle des glaciations (100 000 ans). L'\'etude de Caillon et al. (2003)\cite{caillon_2003} est cit\'ee pour quantifier ce d\'ephasage, de l'ordre de 800 ans. Le CO$_2$ ne d\'eclenche donc pas les d\'eglaciations. Ceci est connu depuis les travaux de Milankovitch du d\'ebut du XX$^e$ siècle\cite{milankovic_1920}. Les climatologues ont expliqu\'e avant l'\'etude de Caillon et al. que le CO$_2$ et les autres gaz \`a effet de serre ne font qu'amplifier les d\'eglaciations qui r\'esultent des for\c{c}cage astronomique d\'ecrits par Milankovitch (cf. Lorius et al. (1990)\cite{lorius_1990}):

\begin{quote}
Changes in the CO$_2$ and CH$_4$ content have played a significant part in the glacial-interglacial climate changes by amplifying, together with the growth and decay of the Northern Hemisphere ice sheets, the relatively weak orbital forcing.
\end{quote}
  
\begin{figure}[ht]
\begin{center}
\includegraphics[width=10cm]{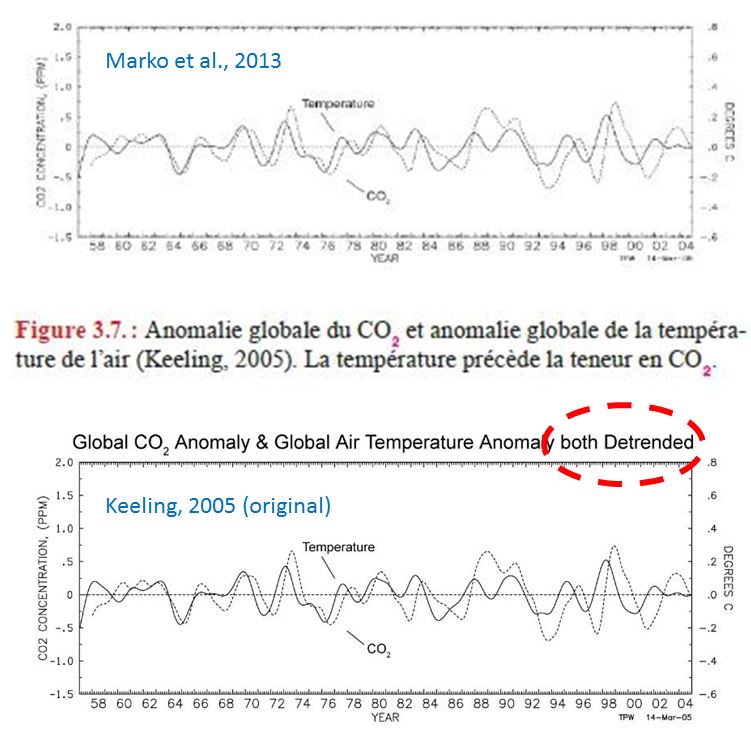}
\end{center}
\caption{En haut: Figure 3.7 de Markó et al. utilis\'ee par les auteurs pour montrer que les variations de CO$_2$ retardent par rapport \`a la temp\'erature. La figure originale, d'apr\`es la source de Markó et al. est reproduite en bas. Curieusement, le terme 'detrended' n'a pas \'et\'e traduit. Il signifie qu'on a retir\'e la composante basse fr\'equence du signal pour en \'etudier les variations rapides. C'est justement les variations lentes de temp\'erature qui définissent le changement climatique (voir figure~\ref{fig:tempglobal}).}
\label{fig:keeling}
\end{figure}

Le deuxi\`eme graphique de Markó et al. (reproduit ici sur la figure~\ref{fig:keeling}) est extrait d'une pr\'esentation de Charles Keeling et montre des variations des anomalies de CO$_2$ et de temp\'erature de 3 \`a 5 ans au XXe si\`ecle. Le CO$_2$ ne d\'eclenche pas non plus les variations de temp\'erature \`a cette \'echelle, qui est celle des oscillations El Ni\~{n}o. A nouveau, ceci \'etait connu des climatologues avant Markó, par exemple dans l'\'etude de Heimann et Reichstein (2008) \cite{heimann_2008}:

\begin{quote}
There is ample empirical evidence that the terrestrial component of
the carbon cycle is responding to climate variations and trends on a
global scale. This is exemplified by the strong interannual variations in
the globally averaged growth rate of atmospheric CO$_2$, which is tightly
correlated with El Ni\~{n}o-Southern Oscillation climate variations.
\end{quote}

Il est bon de rappeler ici au lecteur que la th\'eorie du changement climatique anthropique pr\'edit que le \emph{climat} est influenc\'e par les gaz \`a effet de serre d'origine anthropique.  Le climat est d\'efini par les valeurs moyennes des param\`etres m\'et\'eorologiques sur plusieurs d\'ecennies. 

\clearpage

\begin{figure}[ht]
\begin{center}
\includegraphics[width=10cm]{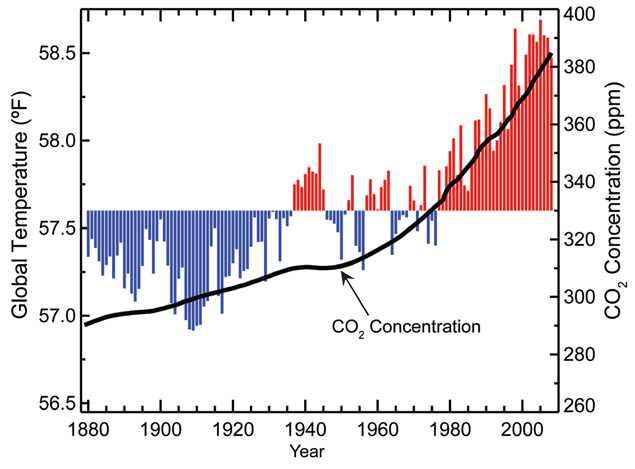}
\end{center}
\caption{Évolution du CO$_2$ et de la temp\'erature globale\cite{global_2009}. L'augmentation de CO$_2$ s'accélère \`a partir de 1950 et la temp\'erature suit à partir des ann\'es 70, contrairement à ce qu'affirme Markó et al. dans la contrev\'erit\'e~\ref{sec:keeling}.}
\label{fig:tempglobal}
\end{figure}

Il y a un problème fondamental dans la reproduction des anomalies de Keeling par Markó et al. La figure de Markó et al. et l'originale de Keeling sont reproduites ici dans la figure~\ref{fig:keeling}. Les courbes sont semblables mais un terme n'a pas \'et\'e traduit de la version de Keeling: 'detrended'. Ce terme signifie que Keeling a soustrait la variation lente du signal pour \'etudier les variations El Ni\~{n}o. C'est justement cette variation lente qui d\'efinit le changement climatique. Il est curieux que Markó et al. aient oubli\'e de traduire ce terme dans leur version de la  l\'egende. Il est aussi curieux qu'ils aient choisi pr\'ecisemment une figure qui ne permette pas \emph{par construction} d'appréhender une relation entre le climat et le CO$_2$. Nous reproduisons une telle figure ici (figure~\ref{fig:tempglobal}, extraite d'un rapport au congr\`es am\'ericain\cite{global_2009}). On peut voir l'augmentation de CO$_2$ s'accélérer \`a partir de 1950 et la temp\'erature suivre à partir des années 70, ce qui r\'efute Markó et al.

\subsection{Contrev\'erit\'e: les volcans \'emettent plus de CO$_2$ que les activit\'es humaines}

\begin{quote}
[I]l y aurait pr\`es de 3,5 millions de volcans sous-marins \ldots \ produisant, \textbf{suivant les diff\'erentes mani\`eres de calculer, des quantit\'es de CO$_2$ allant de 24,5 GtC/an (Kerrick and Caldeira, 1998; Kerrick, 2001)} à 121 GtC/an (Casey, 2010). Ces chiffres sont à mettre en regard de la contribution anthropique reconnue ($\pm$5 à 8 GtC/an). (Markó et al., 64, c'est moi qui souligne)
\end{quote}

Kerrick and Caldeira (1998)\cite{Kerrick1998} est une \'etude sur l'éocène, qui s'étend de 55.8 à 33.9 million avant notre ère, bien avant l'apparition  d'homo sapiens, ce qui n'est pas vraiment pertinent pour comparer avec la contribution anthropique. Casey (2010) est en fait un pseudo-article scientifique publi\'e sur un blog personnel\footnote{\url{http://carbon-budget.geologist-1011.net/}} et ne sera pas consid\'er\'e ici. Kerrick (2001)\cite{kerrick_2001} est par contre une \'etude s\'erieuse sur les \'emissions actuelles des volcans sous-marins. Voici un extrait du r\'esum\'e de l'\'etude: 

\begin{quote}
the total CO$_2$ discharge from subaerial volcanism is estimated at 2.0 –2.5 10$^{12}$ mol/yr.
\end{quote}

Markó et al. pourraient-ils nous expliquer par \textbf{quelle mani\`ere de calculer} ils convertissent cette estimation de Kerrick en 24,5 GtC/an? Une mole de carbone pesant environ 12 g, la valeur maximale estim\'ee par Kerrick (2001) est donc de 0.03 GtC/an, soit 3 ordres de grandeur inf\'erieur à Kerrick (2001) version Markó et al., et en particulier, largement inf\'erieur au chiffre donn\'e par Markó et al. pour la contribution anthropique (5 à 8 GtC/an). 

\subsection{Contrev\'erit\'e: le CO$_2$ \'etait plus abondant au XIX$^e$ siecle qu'aujourd'hui}
\label{sec:beck}

\begin{quote}
\textbf{A titre d'exemple, les teneurs en CO$_2$ mesur\'ees chimiquement (par titrim\'etrie) de 1812 à 1961 (90 000 mesures avec une pr\'ecision de 3\%, souvent meilleure que 1\%) (Beck 2006, Figure 1.5.), dans l'h\'emisphère nord, montrent une forte variabilit\'e avec des teneurs pr\'eindustrielles allant jusqu’à 450 ppm.} (Markó et al., 68, souligné par les auteurs) 
\end{quote}

Curieusement, la r\'ef\'erence Beck (2006) de la bibliographie de Markó et al. (p. 135) n'est pas l'article original de Beck\cite{beck_2007}, mais un r\'esum\'e:

Beck E.G. 2006. 180 years accurate CO2–gas analysis of air by chemical methods (short version).

Voici une note qui figure au bas de chaque page de ce r\'esum\'e:

\begin{quote}
This is an unofficial extract of E-G Beck's comprehensive draft paper and is for discussion not citing
\end{quote}

Pourquoi Markó et al. pr\'eferrent-ils citer le r\'esum\'e plutôt que l'article original, contrairement à l'avertissement de Beck? 

L'article original de Beck a \'et\'e publi\'e dans une revue confidentielle: \emph{Energy and Environment} (facteur d'impact =0.28). Sa lecture est amusante, on peut y lire par exemple:

\begin{quote}
Measurements made prior to 1857 (introduction of Pettenkofer method, 3\% accuracy), mostly by French authors (Boussingault, [14]; Brunner [14]; Regnault [14], [75]), show systematic errors due to long connections (absorption in caoutchouc), H$_2$SO$_4$ for drying air and missing temperature management. There being no calibration against Pettenkofer or modern volumetric/manometric equipment, so I cannot quantify accurately the range of error. \cite{beck_2007} 
\end{quote}

Autrement dit, les mesures avant 1857 ne veulent rien dire puisqu'on ne peut pas estimer leurs incertitudes. En fait, la partie de la courbe après 1857 n'a pas beaucoup plus de sens. Admettons les 3\% de pr\'ecision après 1857 pour la discussion. Cette précision ne s'appliquerait évidemment qu'aux données individuelles des stations de mesures prises en compte par Beck pour la moyenne, pas à la moyenne elle-même. Ces stations sont presque toutes situées près des villes d'après Beck. On sait que le CO$_2$ est très variable près des sources, comme à Paris\cite{widory_2003}. Comment peut-on prétendre montrer une moyenne réaliste pour l'hémisphère Nord avec ces données urbaines? Il est amusant de constater que les mêmes Markó et al. contestent (p. 88) la moyenne des données de température mondiale, beaucoup plus nombreuses et représentatives, pour une raison semblable.   
 
\begin{figure}[ht]
\begin{center}
\includegraphics[width=12cm]{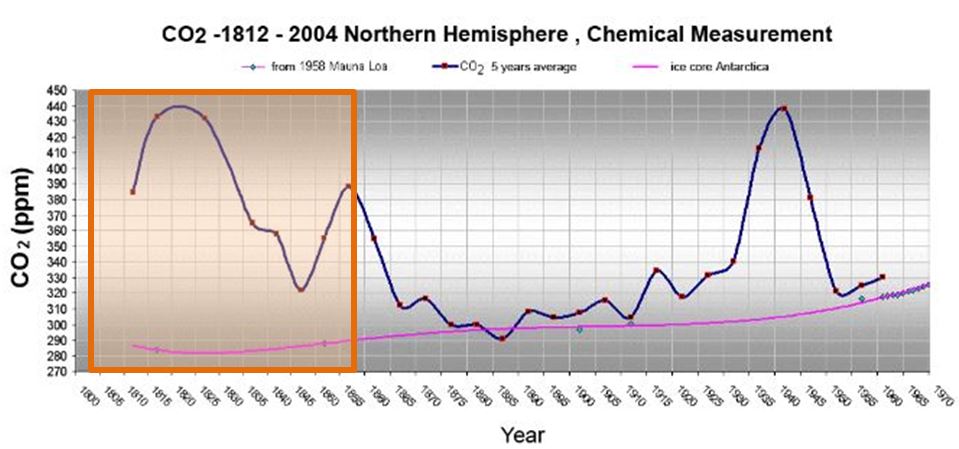}
\end{center}
\caption{La courbe de Beck présentée par Markó et al. qui mentionne 3\% de précision. En fait, Beck ne peut même pas évaluer la précision des données utilisées avant 1857 (zone orange). Les données plus récentes correspondent à quelques stations urbaines où le CO$_2$ est très variable, pour lesquelles il apparaît très périlleux d'extraire une moyenne représentative de l'hémisphère Nord.}
\label{fig:beck}
\end{figure}

\subsection{Contrev\'erit\'e: 450 articles scientifiques climatosceptiques ont \'et\'e publi\'es dans des revues scientifiques prestigieuses}
\label{sec:450}

\begin{quote}
Quoique quelque 450 articles scientifiques climatosceptiques aient \'et\'e publi\'es dans les revues scientifiques les plus prestigieuses (Nature, Science, etc.), il est difficile de publier pour les climatosceptiques. (Markó et al., 49)
\end{quote}

Nous laissons le lecteur apprécier la logique de cette phrase. Pour compter les articles climatosceptiques, il faut au pr\'ealable d\'efinir le terme \emph{climatosceptique}. Voici la d\'efinition de Markó et al.:

\begin{quote}
En r\'ealit\'e, ce qui est contest\'e par les climatosceptiques est l'impact pr\'edominant de l'activit\'e humaine (li\'ee
aux \'emissions des combustibles fossiles) sur le changement climatique. (Markó et al., 51)
\end{quote}

Markó et al. définissent mal les termes qu'ils utilisent. Le changement climatique qui est attribué à l'activité humaine par les climatologues avec un degré de certitude évalué à 95\% par le GIEC est celui observé depuis 1950. Le GIEC ne prétend pas que le changement climatique au moment de l'optimum médiéval est lié à l'utilisation des combustibles fossiles par exemple.  Voici la liste exhaustive des articles de Nature, Nature Geoscience, et Science (facteurs d'impact respectifs pour 2012: 24.5, 9.275, 18.3) qui apparaissent dans la bibliographie de Markó et al.:

\begin{enumerate}

\item  Briffa R., Schweingruber F.H., Jones P.D., Osborn T.J., Shiyatov S.G., Vaganov A. 2008. Reduced sensivity of recent tree-growth to temperature at high northern latitudes. Nature, 391, 678-682.

\item Andresen C.S., Straneo F., Ribergaard M.H. et al. 2012. Rapid response of Helheim Glacier in Greenland climate variability over the past century. Nature Geoscience Letters, 37-41.

\item Kasting J.F. 2010. Faint young Sun redux. Nature, 464(1), 687-688.

\item Schmittner A., Bard E. 2012. Global warming preceded by increasing carbon dioxide concentrations during the last deglaciation, Nature, 484, 49-55.

\item Svensen H., Planke S., Malthe-Sorenssen A., Jamtveit B.,Myklebust R, Eidem T., Rey S.S. 2004. Release of methane from a volcanic basin as a mechanism for initial Eocene global warming. Nature, 429, 542-545.

\item Caillon N., Severinghaus J.P., Jouzel J., Barnola J.M., Kang J., Lipenkov V.Y. 2003. Timing of atmospheric CO2 and Antarctic
temperature changes across Termination III. Science, 299, 5613, 1723-1731

\end{enumerate}

Aucun de ces articles n'est climatosceptique selon la d\'efinition de Markó et al. complétée plus haut. Les \'etudes r\'eellement climatosceptiques sur lesquelles s'appuient l'ouvrage (par exemple, celle qui affirme que le CO$_2$ \'etait plus abondant au XIXe si\`ecle, cf. section~\ref{sec:beck}) ont \'et\'e publi\'e dans des revue comme \emph{Energy and Environment} (facteur d'impact 2012: 0.28). 

\subsection{Contrev\'erit\'e : La temp\'erature et le CO$_2$ \\sont mal corr\'el\'es}
\label{sec:correlation}

\begin{quote}
Aucune corr\'elation entre le r\'echauffement plan\'etaire global et la teneur en CO$_2$ atmosph\'erique ne peut \^{e}tre mise en \'evidence. [\ldots] Akasofu (2010) va plus loin en cherchant une corr\'elation avec le CO$_2$ « fossile » ou anthropique. La figure 3.4., qui provient de l’Institut de M\'et\'eorologie Japonais, montre que la temp\'erature moyenne peut \^etre estim\'ee ou approch\'ee par une relation lin\'eaire en fonction du temps (T = at), mais que les modifications des \'emissions de CO$_2$ anthropique sont, par contre, mieux repr\'esent\'ees par une relation quadratique de type T = at$^2$ (T = temp\'erature, t = temps, a = nombre scalaire.) (Markó et al., 91, 92)
\end{quote}

Markó et al. confondent l'évolution des émissions de CO$_2$ avec celle de sa concentration, c'est cette dernière quantité qui influence l'effet de serre. De plus, Markó et al. n'expliquent pas quelle relation la th\'eorie du r\'echauffement anthropique prévoit entre température et CO$_2$. Cette question \'etait déjà traitée par Arrhenius. Voici ce que ce dernier écrivait en 1897:

\begin{quote}
If the quantity of carbonic acid increases in geometric progression, the augmentation of the temperature will increase nearly in arithmetic progression \cite{arrhenius_1897}.
\end{quote}

Plus d'un siècle après Arrhenius, la physique de l'effet de serre s'est précisée, mais la relation entre température et CO$_2$ a gardé la même forme mathématique. Voici par exemple un extrait du TAR (2001):

\begin{quote}
It has been suggested that the absorption by CO$_2$ is already saturated so that an increase would have no effect. This, however, is not the case. Carbon dioxide absorbs infrared radiation in the middle of its 15 $\mu$m band to the extent that radiation in the middle of this band cannot escape unimpeded: this absorption is saturated. This, however, is not the case for the bands wings. It is because of these effects of partial saturation that the radiative forcing is not proportional to the increase in the carbon dioxide concentration but shows a logarithmic dependence. Every further doubling adds an additional 4 Wm-2 to the radiative forcing.
\end{quote}

Le lecteur intéressé par la physique sur laquelle repose l'effet logarithmique du CO$_2$ sur la température pourra consulter le livre de référence: \emph{An Introduction to Atmospheric Physics}\cite{andrews_2010}.

\begin{figure}[ht]
\begin{center}
\includegraphics[width=12cm]{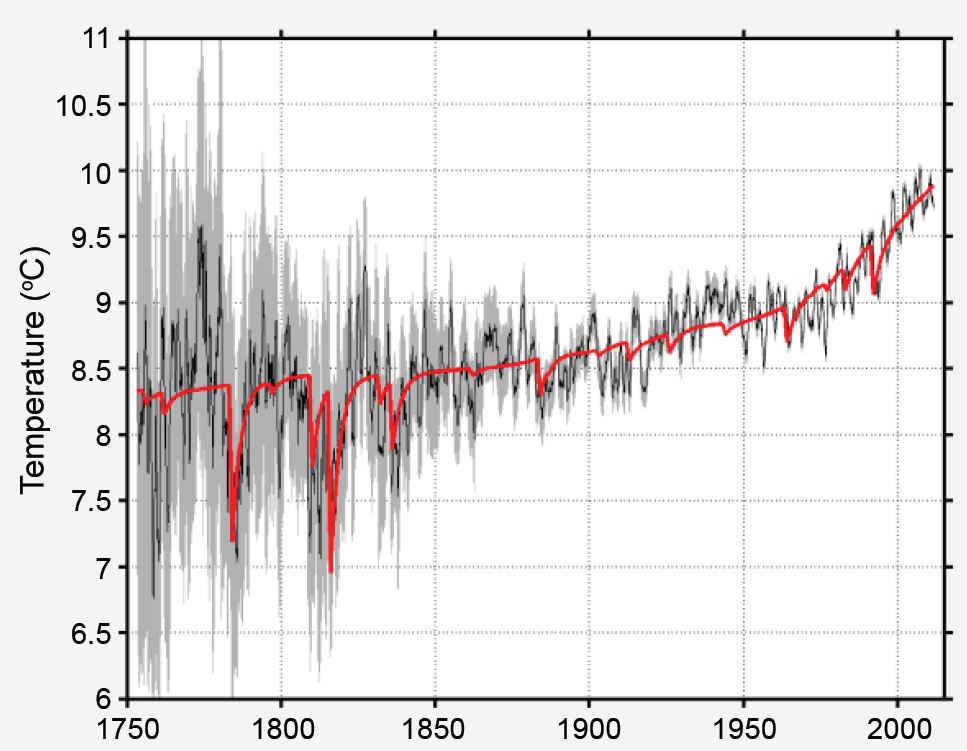}
\end{center}
\caption{The annual [\ldots] land surface temperature from
the Berkeley Earth average (black line), compared to a linear combination of
volcanic sulfate emissions (responsible for the short dips) and the natural
logarithm of CO2 (responsible for the gradual rise) shown in red. Inclusion of a
proxy for solar activity did not significantly improve the fit. Temperature data is
the same as figure 1. The grey area is the 95\% confidence interval.\cite{rohde_2013}}
\label{fig:best}
\end{figure}

La figure~\ref{fig:best} est extraite d'une étude du groupe Berkeley Earth\cite{rohde_2013} (anciennement BEST) et montre l'accord entre un modèle simple prenant en compte l'effet logarithmique du CO$_2$ sur la température. BEST semble être digne de confiance pour Markó et al. qui citent ce groupe à deux reprises (p.80 et p.122) sans le critiquer (sur BEST, voir aussi la section~\ref{sec:best}).

\subsection{Contrev\'erit\'e: 50\% des sp\'ecialistes am\'ericains du climat sont climato-sceptiques}
\label{sec:specialist}

\begin{quote}
Le 19 octobre 2009, le Bulletin de la Soci\'et\'e m\'et\'eorologique am\'ericaine a publi\'e les r\'esultats d’un vote qui a eu lieu parmi les \textbf{sp\'ecialistes am\'ericains du climat} relativement \`a l'origine anthropique du changement climatique : \textbf{50\% des sond\'es ne croient pas \`a l'influence de l'homme sur le climat}, 25\% restent neutres et 8\% croient fermement \`a l'influence de l'homme sur le climat. (Markó et al., 146, soulign\'e par les auteurs)
\end{quote}

Curieusement, ce n'est pas l'article de Bull. Amer. Meteor. Soc. qui est cit\'e par Markó et al., mais sa reprise par le Heartland Institute\footnote{\url{http://news.heartland.org/newspaper-article/2010/02/01/meteorologists-reject-uns-global-warming-claims}}. Il y a bien un article dans Bull. Amer. Meteor. Soc., et les chiffres donn\'es par Markó et al. sont corrects. Voici l'article:

Opportunities and Obstacles for Television Weathercasters to Report on Climate Change, Bull. Amer. Meteor. Soc., 90, 1457–1465.

Les \textbf{sp\'ecialistes am\'ericains du climat} de Markó et al. sont en fait \textbf{des pr\'esentateurs m\'et\'eos à la t\'el\'evision}. 

\subsection{Contrev\'erit\'e: aucun mod\`ele ne conduit \`a des variations cycliques de CO$_2$}

\begin{quote}
Nous constatons qu'aucun des mod\`eles du cycle du carbone
test\'es à ce jour ne conduit \`a des variations cycliques de la teneur
en CO$_2$ de l'atmosph\`ere. (Markó et al, 132)
\end{quote}

\begin{figure}[ht]
\begin{center}
\includegraphics[width=12cm]{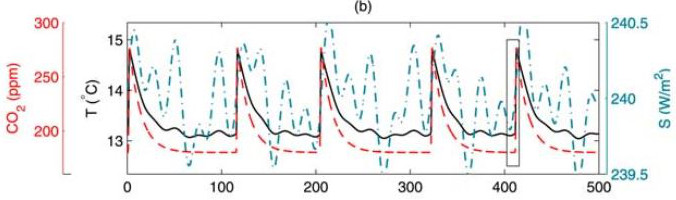}
\end{center}
\caption{Markó et al. affirment qu'aucun mod\`eles du cycle du carbone ne conduit \`a des variations cycliques du CO$_2$ de l'atmosph\`ere. La figure ci-dessus pr\'esente de telles variations calcul\'ees \`a partir d'un mod\`ele simple et pédagogique\cite{hogg_2008}.}
\label{fig:hogg}
\end{figure}

La figure~\ref{fig:hogg} est tir\'ee d'une \'etude (Hogg, 2008)\cite{hogg_2008} qui pr\'esente un modèle simple et pédagogique pour reproduire les variations cycliques de CO$_2$ à l'échelle des glaciations. Seules 7 équations sont nécessaires pour arriver à ce résultat qui contredit Markó et al.

\subsection{Contrev\'erit\'e: Le GIEC réduit ses pr\'edictions de mont\'ee des eaux à chaque rapport}

\begin{quote}
[A] chaque rapport du GIEC, la vitesse de montée des océans est revue \`a la baisse. (Markó et al., 113)
\end{quote}

\begin{sidewaysfigure}[ht]
\begin{center}
\includegraphics[width=18cm]{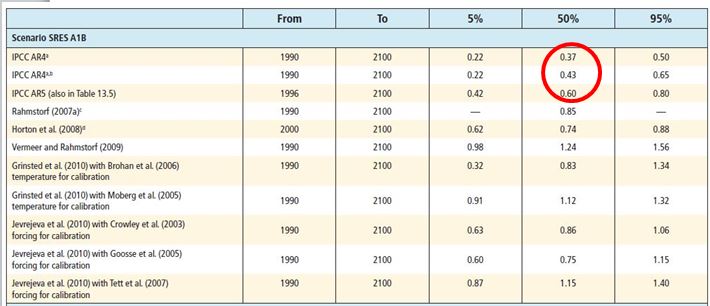}
\end{center}
\caption{Extrait de la table 13.6 de l'AR5. La projection du niveau de mer pour 2100 a augmenté de 50\% par rapport à l'AR5. D'après Markó et al., cette projection diminue à chaque rapport du GIEC. contrairement aux affirmations de Markó et al.}
\label{fig:tab13p6}
\end{sidewaysfigure}

Pour comparer la montée des hauts dans les rapports du GIEC de manière sensée, il faut s'appuyer sur les mêmes scénarios d'émissions de gaz à effet de serre. Cet exercice a été effectué dans l'AR5 du GIEC\cite{IPCC2013}, dont la table 13.6 est montrée ici sur la figure~\ref{fig:tab13p6}. Les deux valeurs médianes de l'AR4 de 2007 sont de 37 et 43 cm, alors que la valeur de l'AR5 de 2013 est de 60 cm. L'AR5 a été publié après Markó et al, mais ceux-ci auraient pu prévoir cette augmentation s'ils s'étaient intéressés la littérature scientifique déjà disponible, comme les autres études de la table 13.6 du GIEC. Celles-ci semblent indiquer que la valeur de 60 cm est conservatrice. 
 
\subsection{Contrev\'erit\'e: la convection est constante dans les mod\`eles du GIEC}

\begin{quote}
[L]es mod\`eles du GIEC ne tiennent pratiquement pas compte des transferts de chaleur convectifs. Ceux-ci sont consid\'er\'es comme constants, donc n'ayant pas d'influence sur le r\'esultat d'un changement par une augmentation du CO$_2$ \og fossile \fg. (Markó et al., 111)
\end{quote}

La convection est variable et influe les résultats de température depuis les premiers modèles. Le lecteur pourra consulter l'ouvrage de référence: \emph{An Introduction to Atmospheric Radiation}\cite{liou_2002} qui décrit justement le traitement de la convection dans ces premiers modèles. Voici un extrait d'un des premiers articles de modélisation de 1975 \cite{manabe_75}, cité par exemple par le GIEC en 1997 \cite{IPCC1997} (c'est moi qui souligne):

\begin{quote}
It should be noted that in low and and middle latitudes, the warming is greater in the upper troposphere [\ldots]. \textbf{This is due to the fact that the moist convective processes in the model tend to adjust temperatures} in a column toward the moist adiabatic lapse rate. 
\end{quote}

\subsection{Contrev\'erit\'e: L'augmentation de la fr\'equence des ouragans est un retour à la normale selon un cycle de 270 ans}

\begin{quote}
MALMGREN BJORN - PhD - Professeur émérite en Géologie
marine et Paléoclimatologie - Göteborg University (S).
« La fréquence des ouragans sur l’Atlantique a été anormalement
faible au cours des années 1970 et 1980. L’augmentation
constatée depuis 1995 est un retour à la normale, selon un
cycle de 270 ans. »
\url{http://www.nature.com/nature/journal/v447/n7145/edsumm/e070607-11.html}
Marko et al., 241
\end{quote}

La citation ci-dessus est extraite de l'annexe II de Markó et al. Cette annexe consiste en une liste de \og scientifiques désapprouvant la thèse de l'origine anthropique du réchauffement climatique \fg (Marko et al., 201), liste dress\'ee par \og \textbf{John ROTHSCHILD} ancien cadre dans le secteur des assurances et animateur du blog \textbf{Belgotopia104} consacr\'e a la probl\'ematique du changement climatique (Markó et al, 201, souligné par les auteurs)\fg. On peut se demander comment un cycle de 270 ans des cyclones pourraient etre connu s'il existait. Pour le savoir, il suffit de cliquer sur l'hyperlien. Contrairement à ce que prétendent Markó et al., celui-ci ne renvoie pas à une citation de Malmgren, mais à un résumé éditorial de \emph{Nature}:

\begin{quote}
The frequency of major hurricanes over the Atlantic Ocean has increased significantly since 1995, but it is still not clear whether this is due to global warming or natural variability. One way to address this question is to consider changes in hurricane frequency in the past, but reliable observations of Atlantic hurricane activity only cover a few decades. Nyberg et al. use proxy records from corals and a marine sediment core that appear to reflect changes in the two main parameters that influence hurricane activity — vertical wind shear and sea surface temperature — to reconstruct the frequency of major hurricanes over the Atlantic since 1730. The results indicate that the frequency was anomalously low during the 1970s and 1980s compared with the past 270 years, and that the phase of increased hurricane frequency since 1995 represents a recovery to 'normal' hurricane activity. These trends appear to be related to wind shear, but what caused this parameter to change remains uncertain.
\end{quote}

Il n'y a pas de cycle de 270 ans, c'est une invention de John Rotschild, comme la citation de Malmgren elle-même. Les 270 ans correspondent à la période sur laquelle nous avons des informations. Notons que Markó et al. précisent avant de présenter la liste que:

\begin{quote}
Les citations figurant dans ce texte ont toutes été vérifiées. (Marko et al., 201)
\end{quote}

Nous laisserons le lecteur réfléchir au fait que malgré les milliers de scientifiques climatosceptiques présents dans ce genre de listes, ces scientifiques n'ont pas été capables de fonder un journal peer-reviewed avec un facteur d'impact significatif.

\subsection{Contrev\'erit\'e: D'apr\`es BEST, l'oscillation atlantique multid\'ecennale contrôle le climat}
\label{sec:best}

\begin{quote}
Pour les États-Unis, suivant l'\'etude BEST (2011), le climat \`a
l'\'echelle d\'ecennale est influenc\'e principalement par la circulation
oc\'eanique atlantique (\og Atlantic Multidecadal Oscillation \fg)
et pacifique (\og Pacific Decadal Oscillation \fg, ou PDO. (Markó et al., 122, la parenthèse ne se referme pas chez les auteurs)
\end{quote}

Le groupe BEST, aujourd'hui appelé Berkeley Earth\footnote{\url{http://berkeleyearth.org/}}, semble une référence sérieuse pour Markó et al. qui le citent aussi pour étayer leurs thèses p.80. En fait, la référence BEST (2011) de Markó et al. n'est pas une étude mais une page internet qui n'existe plus à l'heure d'écrire cette note: \url{http://berkeleyearth.org/findings/}. Il y a cependant un article de BEST qui traite de l'influence des océans sur le climat\cite{muller_2013}. L'article montre une corrélation entre les indices d'oscillation atlantique multidécennale (AMO) et les variations de température globale. Voici un extrait de la conclusion de cette étude:

\begin{quote}
Such changes may be independent responses to a common
forcing (e.g., greenhouse gases); however, it is also possible
that some of the land warming is a direct response to
changes in the AMO region. If the long-term AMO
changes have been driven by greenhouse gases, then the
AMO region may serve as a positive feedback that amplifies
the effect of greenhouse gas forcing over land. On
the other hand, some of the long-term change in the
AMO could be driven by natural variability, e.g., fluctuations
in thermohaline flow. In that case, the human component
of global warming may be somewhat overestimated.
However, in a recent analysis covering more than 250 years
[Rohde et al., 2013b], the long-term temperature changes
were well correlated to a simple model that only contained
information about CO2 and volcanic events. The strong
association between CO2 and the long-term warming argues
against natural variability as a major contributor to the longterm
(century-scale) temperature rise; however, the AMO
and other factors may have contributed significant variability
on shorter (multidecadal) time scales.
\end{quote}

(La référence Rohde et al., 2013b\cite{rohde_2013}, d'où est extraite la figure~\ref{fig:best}, est discutée dans la section{~\ref{sec:correlation},) 

Si Markó et al. ont lu cet étude, ils confondent causalité et corrélation dans leur interprétation de l'étude de BEST. Il est amusant de noter que Markó et al. accusent les \og carbocentristes \fg de faire cette erreur de logique p. 97, dans le passage où Markó et al. modifient une légende pour servir leur cause (section~\ref{sec:keeling}). 

La raison pour laquelle Markó et al. apprécient BEST est que son directeur, le physicien Richard Muller, a mis en doute publiquement les conclusions du GIEC, en particulier sur la courbe de Mann en crosse de hockey (voir section~\ref{sec:mann}), l'importance de l'effet d'îlot urbain, ou l'attribution de la hausse de température aux émissions anthropiques. Muller expliquait ainsi sa démarche à la \emph{National Public Radio} le 11 avril 2011\footnote{http://www.npr.org/2011/04/11/135320209/climate-change-skeptic-says-warming-is-real}: 

\begin{quote}
Temperature has been rising over the last 100 years. That's pretty clear. How much is due to varying solar activity and how much due to humans is a scientific issue that we're trying to address.  [\ldots]	 \ I have created a new project here in Berkeley - we call it Berkeley Earth - that is doing a reexamination of the global warming issue. We are addressing all of the issues that have been raised - all the legitimate issues that have been raised by the people called the skeptics. And there are some legitimate issues there. 
\end{quote}

Voici maintenant ce qu'écrit Richard Muller dans le New York Times du 28 juillet 2012\cite{muller_2012}:  

\begin{quote}
Call me a converted skeptic. Three years ago I identified problems in previous climate studies that, in my mind, threw doubt on the very existence of global warming. Last year, following an intensive research effort involving a dozen scientists, I concluded that global warming was real and that the prior estimates of the rate of warming were correct. I’m now going a step further: Humans are almost entirely the cause. 
\end{quote}


\section{Remarques sur l'\'epist\'emologie de Markó et al.}
\label{sec:epistem}

\begin{quote}

Long experience has taught me this about the status of mankind with regard to matters requiring thought: the less people know and understand about them, the more positively they attempt to argue concerning them, while on the other hand to know and understand a multitude of things renders men cautious in passing judgment upon anything new.

Galileo Galilei, The Assayer, 1623 \cite{galilei_1957}
\end{quote}

\subsection{La science doit-elle être officielle?}
\label{sec:certif}

\begin{quote}
Une th\'eorie dont on pr\'etend qu'elle ne peut plus \^etre scientifiquement d\'ebattue perd ipso facto son caract\`ere scientifique pour devenir un ensemble de croyances dogmatiques. (Markó et al., 47) 
\end{quote}

\begin{quote}
Le site du gouvernement belge relaye, sans recul critique, les th\`eses du GIEC, leur conf\'erant ainsi une sorte de l\'egitimit\'e institutionnelle. Il est d\'ej\`a critiquable que les services de l'Etat pr\'esentent comme faisant partie de la science officielle des th\`eses qui sont encore discut\'ees dans la communaut\'e scientifique, mais il est encore plus critiquable de ne pas restituer fid\`element ces th\`eses. (Markó et al, 169)
\end{quote}

Il y a une contradiction logique ici. Dans la première citation, les auteurs disent que les propositions scientifiques doivent être soumises au d\'ebat et dans la deuxième que la climatologie du GIEC n’a pas le statut de science officielle puisqu’elle est encore discut\'ee. Que signifie "science officielle" ? Y a-t-il un organisme qui officialise les chercheurs ou les articles? D'un côt\'e, les auteurs critiquent le GIEC parce qu’ils affirment qu'il pr\'etend être un tel organisme de certification et que ça n’existe pas en science, de l'autre ils invoquent une science \og officielle \fg.

\subsection{L'art de mal poser les problèmes}

\begin{quote}
Imaginons un instant que nous nous soyons tromp\'es \ldots	 \ qu'en d\'epit de tous les arguments et analyses contenus dans ce livre, les th\`èses du GIEC soient n\'eanmoins la v\'erit\'e une et d\'efinitive en mati\`ere climatique. (Markó et al, 182-183, suivi d'une discussion sur l'intérêt du débat et de la liberté d'expression, puis une longue citation de John Stuart Mill)  
\end{quote}

La dichotomie de Markó et al. est irréaliste et contraire à l’esprit scientifique. Les thèses de Markó et al. peuvent bien être complètement délirantes et les rapports du GIEC contenir simultanément des erreurs. Qui a jamais prétendu que les thèses du GIEC soient « la vérité une et définitive» ? Dans l’AR5, la responsabilité humaine est décrite comme établie à 95\%. Manquent donc 5\% pour une « vérité une et définitive », qui n’existe en fait jamais en science. Les rédacteurs du GIEC seraient-ils en fait tous climatosceptiques ?

\subsection{Sur Galil\'ee, Von Neumann, et Feynman}

Deux g\'enies scientifiques apparaissent pour étayer les \og thèses \fg de Markó et al.: Galilée (1564-1642) et John Von Neumann (1903-1957). Leur utilisation prête à sourire.

\begin{quote}
Le consensus n’est pas un critère de vérité. Il existait un \og consensus \fg scientifique contre les thèses de Galilée au XVIe siècle. (Markó et al., 47)
\end{quote}

Invoquer Galilée pour impressionner la galerie et justifier des opinions farfelues est on ne peut plus banal. Mais pour rappel, ce n’était pas avec les scientifiques que Galilée avait un problème, c’est avec l’Eglise (voir \emph{Religion and Science}\cite{russell_1997}, de Bertrand Russell). La science moderne ayant commencé avec Galilée, on ne voit pas trop comment il aurait pu y avoir un \og consensus \fg scientifique contre ses thèses. Les quelques représentants de la science à l’époque, des gens comme Kepler ou Descartes, étaient avec Galilée contre l’Eglise. 

\begin{quote}
Comme le faisait remarquer un mathématicien célèbre, John von Neuman \og avec quatre paramètres je vous dessine un éléphant; si vous m’en donnez un cinquième, je m’arrangerai pour qu’il remue la trompe \fg. (Markó et al., 100, les auteurs n'écrivent pas correctement le nom du mathématicien célèbre)
\end{quote}

John Von Neumann (1903-1957) a fait beaucoup de choses. Il est amusant de le citer pour critiquer les modèles alors qu’il est le père des modèles météo\cite{Heymann-2010}. Par ailleurs, Von Neuman s’est exprimé sur le réchauffement du au CO$_2$ anthropique dès 1955\cite{vonNeumann-1955}, soit plus de 30 ans avant la naissance du GIEC: 

\begin{quote}
The carbon dioxide released into the atmosphere by industry's burning of coal and oil--more than half of it during the last generation--may have changed the atmosphere's composition sufficiently to account for a general warming of the world by about one degree Fahrenheit.
\end{quote}

La lecture de Markó et al. fait penser à un texte magnifique de Richard Feynman (1918-1988), \emph{Cargo Cult Science}\cite{feynman_1974}, qui devrait faire partie de la culture générale de tout scientifique. Voici un extrait de \emph{Cargo cult science} (c'est moi qui souligne): 

\begin{quote}
I would like to add something that's not essential to the science, but something I kind of believe, which is that \textbf{you should not fool the layman when you're talking as a scientist}. I am not trying to tell you what to do about cheating on your wife, or fooling your girlfriend, or something like that, when you're not trying to be a scientist, but just trying to be an ordinary human being. We'll leave those problems up to you and your rabbi. \textbf{I'm talking about a specific, extra type of integrity that is not lying, but bending over backwards to show how you're maybe wrong, that you ought to have when acting as a scientist. And this is our responsibility as scientists, certainly to other scientists, and I think to laymen.} 
\end{quote}

\clearpage

\begin{figure}[ht]
\begin{center}
\includegraphics[width=10cm]{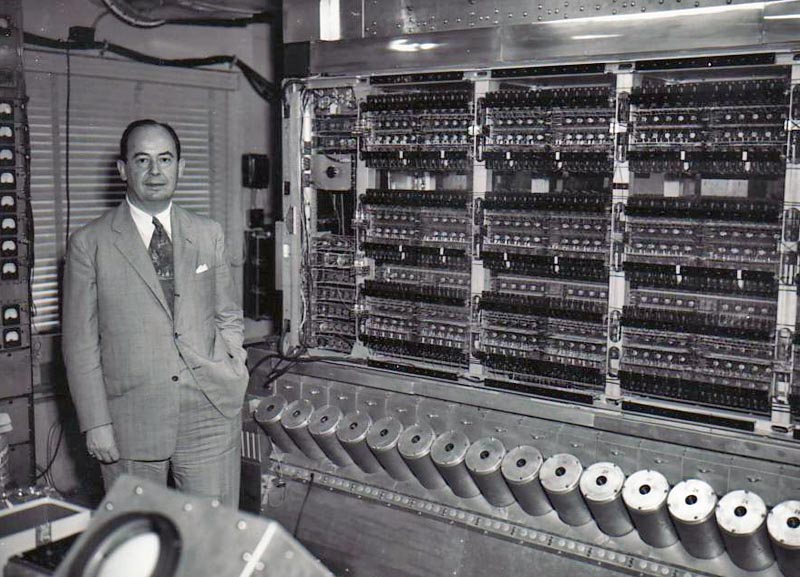}
\end{center}
\caption{John Von Neumann est cité par Markó et al. pour critiquer les mod\`eles. Sur cette photo (Institute for Advanced Study), Von Neumann  est repr\'esent\'e avec l'ordinateur ENIAC qu'il a utilis\'e pour d\'evelopper le premier modèle de pr\'evison meteo. Von Neumann s'est par ailleurs exprim\'e sur le r\'echauffement climatique du au CO$_2$ anthropique dès 1955, plus de 30 ans avant la création du GIEC.}
\label{fig:vonneumann}
\end{figure}

\section{Les sources de Markó et al.}
\label{sec:sources}

Nous avons vu que Markó et al. prétendent s'appuyer sur des études publiées dans des revues scientifiques prestigieuses comme \emph{Nature} (section~\ref{sec:450}) ou sur l'opinion de la moitié des spécialistes américains du climat (section~\ref{sec:specialist}), mais qu'il n'en est rien. L'étude de la bibliographie et des sources de Markó et al. révèle cependant des régularités intéressantes. Comme le montre la figure~\ref{fig:heartland}, un petit nombre d'auteurs et leur site internet sont largement cités: Fred Singer, Richard Lindzen, Anthony Watts, et Roy Spencer. Ces personnes se sont toutes déclarées expertes\footnote{\url{http://heartland.org/richard-lindzen},\url{http://heartland.org/roy-spencer},\url{http://heartland.org/s-fred-singer},\url{http://heartland.org/anthony-watts}} pour un organsisme qui apparait lui aussi à de nombreuses reprises chez Markó et al.: le Hearltand Institute. On notera également que deux des coauteurs de Markó et al. (Anne Debeil et Lars Myren) sont également auteurs pour ce même Heartland Institute\cite{myren_2009}. 

\begin{figure}[H]
\begin{center}
\includegraphics[width=10cm]{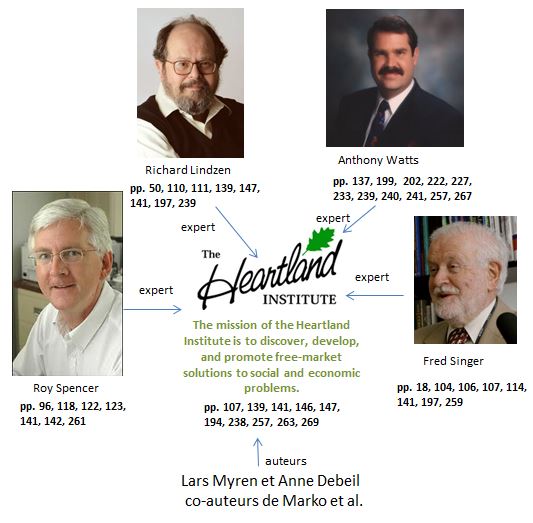}
\end{center}
\caption{Occurences du Heartland Institute et de ses experts dans Markó et al. }
\label{fig:heartland}
\end{figure}

Voici ce qu'écrit \emph{Nature}, revue sur laquelle Markó et al. prétendent s'appuyer, à propos de la climatologie du Heartland Institute: 

\begin{quote}
[\ldots] Nature is not endorsing the Heartland Institute as a serious voice on climate science [\ldots] It is scientists, not sceptics, who are most willing to consider explanations that conflict with their own. And far from quashing dissent, it is the scientists, not the sceptics, who do most to acknowledge gaps in their studies and point out the limitations of their data — which is where sceptics get much of the mud they fling at the scientists. By contrast, the Heartland Institute and its ilk are not trying to build a theory of anything. They have set the bar much lower, and are happy muddying the waters. \cite{nature_2011}
\end{quote}

Il est étrange que les auteurs ne présentent pas le Heartland Institute en expliquant pourquoi il serait une source plus digne de confiance que \emph{Nature} en ce qui concerne la science du climat. 

Le Heartland Institute est-il un groupe de recherche sur le climat? Voici comment le Heartland Institute se définit lui-même\footnote{\url{http://heartland.org/about}}:

\begin{quote}
The Heartland Institute is a 30-year-old national nonprofit research organization dedicated to finding and promoting ideas that empower people[\ldots]Its mission is to discover, develop, and promote free-market solutions to social and economic problems.
\end{quote}

Il apparait donc que la spécialité du Heartland Institute n'est pas la climatologie mais l'économie, et qu'au sujet de cette dernière, le Heartland Institute défend des solutions de libre marché. On notera aussi que Ludovic Delory\footnote{\url{http://www.wikiberal.org/wiki/Ludovic_Delory}} et Drieu Godefridi\footnote{\url{http://www.wikiberal.org/wiki/Drieu_Godefridi}} se définissent tous les deux comme des \og libéraux \fg. 

Voici la dernière des \og vérités qui dérangent \fg de Markó et al.:

\begin{quote}
Les milieux économiques et financiers ont été contraints de s’adapter aux politiques de lutte contre le changement climatique qui affectent durement leur productivité et leur compétitivité alors que d’aucuns ont réussi à se ménager des rentes de situation au détriment de leurs concurrents et des contribuables.
[\ldots]
La littérature économique anglo-saxonne révèle que, depuis quelques années, de grandes entreprises et groupes financiers ont abusé de cet enthousiasme pour les investissements verts afin de se ménager des rentes de situation au détriment des contribuables et/ou des consommateurs. (Markó et al., 175)
\end{quote}

La littérature économique anglo-saxonne? Les deux uniques références de cette section renvoient à des blogs francophones qui défendent eux aussi le libre marché (Institut Turgot\footnote{\url{http://blog.turgot.org/}} et Objectif liberté\footnote{\url{http://www.objectifliberte.fr/}}). 

A ce stade de l'analyse de Markó et al., il semble clair que cet ouvrage s'inscrit dans le combat obscurantiste menée par les partisans du marché libre et décrite par Oreskes et Conway dans \emph{Les marchands de doute}\cite{oreskes_2012}. Les  \og libéraux \fg semblent avoir un réel problème avec la théorie du chagnement climatique. Ainsi, une étude récente publiée dans \emph{Psychological Science}\cite{lewandowsky_2013} et conduite sur plus d'un milliers de blogeurs du climat indique une forte corrélation ($r \simeq  0.8$) entre les partisans du laissez-faire et le rejet de la science du climat.  

Voyons maintenant les contradictions entre théorie de l'efficacité des marchés libres et théorie du réchauffement climatique anthropique.

\begin{figure}[H]
\begin{center}
\includegraphics[width=10cm]{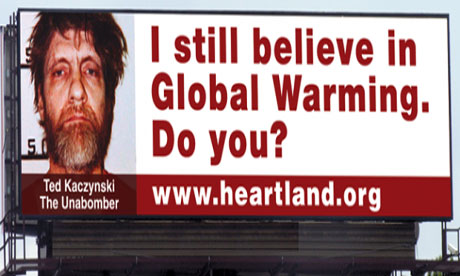}
\end{center}
\caption{Affiche du Heartland Institute à Chicago (Heartland Institute). Extrait du communiqué de presse présentant l'opération : \og Billboards in Chicago paid for by The Heartland Institute point out that some of the world's most notorious criminals say they "still believe in global warming" – and ask viewers if they do, too…The billboard series features Ted Kaczynski, the infamous Unabomber; Charles Manson, a mass murderer; and Fidel Castro, a tyrant. Other global warming alarmists who may appear on future billboards include Osama bin Laden and James J. Lee (who took hostages inside the headquarters of the Discovery Channel in 2010).
These rogues and villains were chosen because they made public statements about how man-made global warming is a crisis and how mankind must take immediate and drastic actions to stop it. \fg Cette campagne a couté très cher au Heartland Institute car elle a entrainé le retrait de plusieurs de ses donateurs et experts. Le Heartland ayant la coutume stalinienne d'effacer les histoires qui lui déplaisent, nous sommes obligés ici de citer une source secondaire (\emph{The Guardian}\cite{hickman_2012}).}
\label{fig:heartland2}
\end{figure}


\section{Parlons d'externalités négatives, voulez-vous?}
\label{sec:externalites}

\begin{quote}
Climate change is a result of the greatest market failure the world has seen. \\
\bigskip
Nicholas Stern, \emph{The Guardian}, 2007 \cite{stern_2007}
\end{quote}

La théorie de l'efficacité du marché libre est supposée remonter à des penseurs des Lumières comme  Anne-Robert-Jacques Turgot(1727-1781) ou Adam Smith(1723-1790)\footnote{Le titre de cette section est empruntée à \href{http://voir.ca/chroniques/prise-de-tete/2012/09/12/parlons-dexternalites-negatives-voulez-vous/}{une chronique de Normand Baillargeon publiée par \emph{voir.ca}}}. Ce dernier, père de la pensée économie moderne, est particulièrement cité pour sa métaphore de la \emph{main invisible}.  Ainsi, Walter Block, dans son article \emph{Defending the speculators}, publié initialement par le Von Mises Institute\cite{Block_2006}  (lui aussi dédié à défendre les bienfaits du marché libre) et repris par le Heartland Institute\footnote{\url{http://heartland.org/policy-documents/defending-speculators}}, écrit: 

\begin{quote}
According to this doctrine, “every individual endeavors to employ his capital so that its produce may be of the greatest value. He generally neither intends to promote the public interest, nor knows how much he is promoting it. And he intends only his own security, his own gain. He is led in this as if by an invisible hand to promote an end that was no part of his intention. By pursuing his own interest he frequently promotes that of society more effectually than when he really intends to promote it.”[i]
\end{quote}

La référence [i] de Block renvoie à l'ouvrage majeur de Smith, publié en 1776, \emph{An Inquiry into the Nature and Causes of the Wealth of Nations}\cite{smith_1904} mais il est précisé par Block que la citation a été \og paraphrased\fg. Comme la plupart des commentateurs \og libéraux \fg, Block ne cite pas le début de cette phrase, la seule de \emph{Wealth of Nations} où est mentionnée la main invisible: \emph{By preferring the support of domestic to that of foreign industry}. Le lecteur qui s'intéresserait à la pensée d'Adam Smith est invité à le lire dans le texte. La main invisible d'Adam Smith est analysée, dans son contexte et de manière lumineuse,  par l'économiste contemporain Michael Meeropol, dans son article \emph{Another distortion of Adam Smith: The case of the “invisible hand.” }\cite{meeropol_2004}. Pour une autre discussion, en français, de l'orthogonalité de ce qu'on appelle \og les politiques néolibérales \fg avec la pensée et les écrits d'Adam Smith, le lecteur est invité à lire Francisco Vergara, \emph{Les fondements philosophiques du libéralisme}\cite{vergara_2002}. 

Plus de deux siècles après Adam Smith, la théorie économique a-t-elle fait des progrès? Voici ce que dit Joseph Stiglitz, \og prix Nobel d'Économie \fg (2001) \cite{stiglitz_2006} (c'est moi qui souligne):
   
\begin{quote}
Adam Smith, the father of modern economics, is often cited as arguing for the "invisible hand" and free markets: firms, in the pursuit of profits, are led, as if by an invisible hand, to do what is best for the world. But unlike his followers, Adam Smith was aware of some of the limitations of free markets, and research since then has further clarified why free markets, by themselves, often do not lead to what is best [\ldots] \textbf{Whenever there are "externalities"---where the actions of an individual have impacts on others for which they do not pay, or for which they are not compensated---markets will not work well}. Some of the important instances have long understood environmental externalities. \textbf{Markets, by themselves, produce too much pollution. Markets, by themselves, also produce too little basic research}. (The government was responsible for financing most of the important scientific breakthroughs, including the internet and the first telegraph line, and many bio-tech advances.) But recent research has shown that these externalities are pervasive, whenever there is imperfect information or imperfect risk markets---that is always. 
\end{quote}

Le concept d'\emph{externalité} est essentiel pour comprendre l'attaque contre la science menée par le Heartland Institute et autres \og libéraux \fg. Une externalité se produit lorsque la consommation ou la production d'un agent a une influence sur le bien-être d'un autre agent, sans que cette interaction ne fasse l'objet d'une transaction économique. Le concept d'externalité a été introduit dans la théorie économique par Arthur Cecil Pigou (1877– 1959), dans son livre \emph{The Economics of Welfare}\cite{pigou_1920}. 

Même les économistes les plus partisans du laissez-faire reconnaissent l'inefficacité du marché à résoudre les problèmes posés par certaines externalités. Friedrich Hayek (1899-1992), qui avait lui-aussi reçu un \og prix Nobel d'économie \fg (1974), semble être apprécié par certains des auteurs de Markó et al. puisque l'un deux (Drieu Godefridi) a fondé un \og Institut Hayek \fg \footnote{\url{http://www.wikiberal.org/wiki/Drieu_Godefridi}}. Voici ce qu'écrivait Hayek sur les externalités dans \emph{The Road to Serfdom}(1944)\cite{hayek_2001} (c'est moi qui souligne):

\begin{quote}
\textbf{There are, finally, undoubted fields where no legal arrangements
can create the main condition on which the usefulness
of the system of competition and private property depends}:
namely, that the owner benefits from all the useful services
rendered by his property and suffers for all the damages caused 
to others by its use. Where, for example, it is impracticable to
make the enjoyment of certain services dependent on the payment
of a price, competition will not produce the services; and
the price system becomes similarly ineffective when the damage
caused to others by certain uses of property cannot be effectively
charged to the owner of that property. In all these instances there
is a divergence between the items which enter into private calculation
and those which affect social welfare; and whenever this
divergence becomes important some method other than competition
may have to be found to supply the services in question.
Thus neither the provision of signposts on the roads, nor, in
most circumstances, that of the roads themselves, can be paid for
by every individual user. \textbf{Nor can certain harmful effects of
deforestation, or of some methods of farming, or of the smoke
and noise of factories, be confined to the owner of the property
in question or to those who are willing to submit to the damage
for an agreed compensation. In such instances we must find
some substitute for the regulation by the price mechanism.} 
\end{quote}

Le principal apport théorique des économistes partisans du marché libre postérieurs à Hayek sur les externalités a été fourni par Ronald Coase (1910-2013)\og prix Nobel d'économie \fg (1974). C'est en particulier l'article que Coase a publié en 1960 \emph{The problem of Social Cost}\cite{coase_1960}, qui lui a valu son prix Nobel. Cet article a inspiré le célèbre \og théorème de Coase \fg, qui semble sauver la main invisible des externalités. La première apparition du théorème de Coase dans la littérature économique n'est en fait pas due à Coase mais à un autre \og prix Nobel d'économie \fg (1982), George Stigler, dans son livre \emph{The Theory of Price}\cite{stigler_1987}. Voici ce que Stigler y écrit:

\begin{quote}
The Coase theorem thus asserts that under perfect competition private and social costs will be equal. The proposition that the composition of output will not be affected by the manner in which the law assigns liability for damage seems astonishing. But it should not be. Laws often prove to be unimportant.
\end{quote}

En français et en clair: dans des conditions de marché parfaites, les externalités feraient, comme le reste, l'objet de négociations entre les différentes parties, et ces négociations entraineraient naturellement une solution acceptable pour tous. 

La première chose à savoir sur le théorème de Coase, c'est qu'il n'existe pas, d'après Coase lui-même\footnote{\url{https://www.youtube.com/watch?v=04zFygmeCUA}}. Coase s'opposait à la tendance des économistes à cacher leur irréalisme derrière des mathématiques. Parmi les conditions de marché parfait, il y a l'information parfaite des agents et des coûts de transaction nuls. Ni l'une ni l'autre de ces conditions ne correspondent au monde réel. Dans son article de 1960, Coase attire en particulier l'attention du lecteur sur les coûts de transaction dans le cas précis de la pollution, précisément dans son article \emph{The problem of Social Cost}:

\begin{quote}
there is no reason why, on occasion, such governmental administrative regulation should
not lead to an improvement in economic efficiency. This would seem particularly
likely when, as is normally the case with the smoke nuisance, a large
number of people are involved and in which therefore the costs of handling
the problem through the market or the firm may be high.
\end{quote}

Autrement dit, Coase pensait que, dans le cas de la pollution, trop de gens sont impliqués pour que le marché puisse résoudre le problème efficacement.

Une autre condition du marché parfait pour Coase est l'existence de droits de propriété clairement définis. Ceci pousse les économistes défenseurs du marché libre à demander l'attribution de droits de propriété individuels (c'est à dire la privatisation) pour les parcs, les forêts, les prairies, les routes, les rivières. 	C'est ce qui est suggéré dans la littérature économique \og libérale \fg récente, comme dans cet article de 2009 au titre provocateur: \emph{The End of the Externality Revolution}\cite{Barnett_2009}, qui insiste sur les implications de l'apport théorique de Coase. Mais même pour les auteurs de cette étude, définir des droits de propriété ne permet pas de régler tous les problèmes. Voici un extrait de \emph{The End of the Externality Revolution} (c'est moi qui souligne): 

\begin{quote}
\textbf{With a public good, any individual affected by the public good gets benefits without paying, since there is, by definition, no way to fence out nonpaying parties}. In this case, \textbf{free-rider problems make it costly for private parties to define property rights} such that all impacted parties are included in (Pareto relevant) bargains. This complication may make transactions cost sufficiently high that they swamp even large potential gains from trade. Examples where this is the case are surprisingly difficult to find. \textbf{Air quality, ozone depletion and climate change are possible examples}.
\end{quote}

A nouveau, les économistes les plus fanatiquement partisans du laissez-faire butent sur l'incapacité de ce dernier à gérer les problèmes environnementaux provenant de l'activité économique. On comprend alors mieux l'empressement des marchands de doute (voir section~\ref{sec:intro}) à biaiser l'information du public sur ces questions.  

En fait, le travail de Coase a servi en partie de base théorique à la création des marchés du carbone\cite{brohe_2008}, qui visent à réduire les coûts de transaction en créant un marché officiel encadré et à définir des émissions maximales acceptables par la société, c'est-à-dire à choisir un optimum entre le droits de l'industrie à émettre de gaz à effet de serre et le droit de la société à préserver un environnement vivable. Quelles que soient les critiques que peuvent inspirer cette solution\cite{bernier_2008}, il est évident qu'elle implique une intervention massive de la puissance publique pour établir les plafonds d'émission, organiser et réguler un marché artificiel. Comme l'écrit Beatrice Quenault, \og dans ce cas précis, contrairement à l'adage, plus de marché signifie non pas moins, mais plus d'État. \fg\cite{quenault_2009}.
    
Pour le Heartland Institute et autres fanatiques \og libéraux \fg, il ne peut pas y avoir d'externalité négative de l'ampleur d'un changement climatique global, puisque son existence remet en cause leur version de la main invisible. La climatologie du Heartland Institute évoque un épisode de l'histoire des sciences, bien résumé par un expert du Heartland Institute lui-même, Richard Lindzen\cite{lindzen_2013}:

\begin{quote}
In the Soviet Union, Trofim Denisovich Lysenko (1898-
1976) promoted the Lamarckian view of the inheritance of
acquired characteristics. It fit with Stalin’s megalomaniacal
insistence on the ability of society to remold nature. Under
Communism, the state was its own advocacy organization.
However, opposition within the Soviet Union remained
strong, despite ruthless attempts to suppress dissenters,
and was consistently supported by scientists outside of the
Soviet Union. Eventually, it was able to assert itself after
Stalin’s death. But even then, Lysenko and his supporters
continued in their formal positions. This may have facilitated
ending the dominance of Lysenko since they weren’t
defending their jobs. \cite{lindzen_2013}
\end{quote}

Comme les biologistes partisans des théories de Lyssenko, les scientifiques qui promeuvent la climatologie \og libérale \fg corrompent la science pour défendre une idéologie sur la société. Comme toute analogie, le rapprochement avec Lyssenko a ses limites. Les biologistes de Staline avaient l'excuse de vivre sous un régime politique dictatorial.

.

\section{Conclusion}

\begin{quote}
Rien n’est plus répréhensible à mes yeux que cette disposition à fuir,
cette désertion si caractéristique d’une position de principe difficile
dont on sait pertinemment qu’elle est juste. Cette peur de paraître
trop politique et revendicatif, ce besoin d’approbation de la part
d’un tenant de l’autorité ; ce désir de maintenir une réputation d’objectivité
et de modération dans l’espoir d’être sollicité, consulté ou de siéger
dans quelque comité prestigieux, afin de se maintenir au sein du courant dominant,
et de recevoir peut-être un jour un diplôme, un prix, une ambassade. 

Edward Saïd, \emph{Des Intellectuels et du Pouvoir}, 1994, cité par Serge Halimi, \emph{Le Grand Bond en Arrière}\cite{halimi_2004}. 
\end{quote}

Le livre \emph{Climat: 15 v\'erit\'es qui d\'erangent} contient trop d'erreurs pour être pris au sérieux par un public scientifique, ce qui n'est d'ailleurs sans doute pas son but. Cependant, ce livre est intéressant à plus d'un titre. C'est d'abord un exemple de ce que Richard Feynamn appelait \emph{Cargo Cult Science}. L'ouvrage, qui prétend s'appuyer sur de la littérature scientifique classique, contient nombres de graphiques et de références qui lui confèrent un aspect académique. Mais les propositions révolutionnaires de Markó et al. ne résistent pas à l'épreuve de la vérification des sources. Des trois d\'ebats mentionnés en introduction, ce livre s'inscrit assez clairement dans le premier, celui des \emph{marchands de doute}. La critique de ce livre permet alors d'attirer l'attention du public francophone sur une organisation particulièrement active dans l'obscurantisme à l'égard de la science du climat: le Heartland Institute, et plus généralement sur beaucoup de ceux qu'on appelle étrangement aujourd'hui \og libéraux \fg, la liberté dont il est question pour ces gens étant d'abord celle du marché. Cet obscurantisme s'explique par l'incapacité du marché libre à résoudre une externalité de l'ampleur d'un changement climatique anthropique, qui risque d'entraîner l'espèce humaine dans des conditions climatiques inconnues depuis le début de la civilisation au néolithique.  Comme le font remarquer Oreskes et Conway dans \emph{Les marchands de doute}, l'obscurantisme des \og libéraux \fg sur le climat démontre un peu plus l'irréalisme de la \og théorie \fg de l'efficacité du marché libre puisque celle-ci suppose l'information parfaite des agents. En truquant l'information du public sur l'état de la science du climat, les \og libéraux \fg s'opposent un peu plus au bon fonctionnement du marché. 

Malheureusement, cette doctrine, la \og théorie \fg de l'efficacité du marché libre, bien qu'elle puisse paraître ridicule telle que décrite par les \og libéraux \fg du Heartland Institute, est en fait largement partagée par les décideurs politiques occidentaux, y compris ceux qui prétendent prendre au sérieux la théorie du réchauffement climatique anthropique. Ces décideurs négocient, à l'heure d'écrire cette note et dans la plus grande discrétion, la création d'un grand marché transatlantique. Celui-ci dépouillera un peu plus les habitants de cette planète du seul moyen d'action efficace pour maîtriser les menaces environnementales, et plus généralement pour contrôler leur propre vie: la maîtrise de l'économie. Comme pendant l'affaire Lyssenko du temps de l'Union soviétique, une théorie scientifique sur la nature n'est pas compatible avec les conceptions sur la société de ceux qu'Adam Smith appelait \og les maîtres de l'espèce humaine \fg. L'Église Catholique a mis près de quatre siècles avant d'admettre son erreur sur Galilée et l'héliocentrisme. Le temps que régnera encore la théorie de l'efficacité des marchés libre dépend de plusieurs facteurs. Il est important de rappeler ici au lecteur que l'un de ces facteurs est l'implication des scientifiques conscients dans les débats économiques.  

A propos de ce dernier point, et comme il est apparu que l'ouvrage de Markó et al. était largement destiné à dissimuler de la mauvaise économie derrière de la mauvaise physique, rappelons ici les mots d'un physicien, Albert Einstein (1879-1955), prix Nobel (sans guillemet), lorsque ce dernier s'exprimait ouvertement sur les affaires économiques\cite{Einstein_1949}:

\begin{quote}
[W]e should be on our guard not to overestimate science and scientific methods when it is a question of human problems; and we should not assume that experts are the only ones who have a right to express themselves on questions affecting the organization of society.
\end{quote}

Dans ce même article, Albert Einstein exprimait clairement son scepticisme quant à la théorie de l'efficacité du marché libre, pour des raisons complètement indépendantes des externalités environnementales.

\begin{quote}
Production is carried on for profit, not for use. There is no provision that all those able and willing to work will always be in a position to find employment; an “army of unemployed” almost always exists. The worker is constantly in fear of losing his job. Since unemployed and poorly paid workers do not provide a profitable market, the production of consumers’ goods is restricted, and great hardship is the consequence. Technological progress frequently results in more unemployment rather than in an easing of the burden of work for all. The profit motive, in conjunction with competition among capitalists, is responsible for an instability in the accumulation and utilization of capital which leads to increasingly severe depressions. Unlimited competition leads to a huge waste of labor, and to that crippling of the social consciousness of individuals which I mentioned before.

\begin{figure}[ht]
\begin{center}
\includegraphics[width=8cm]{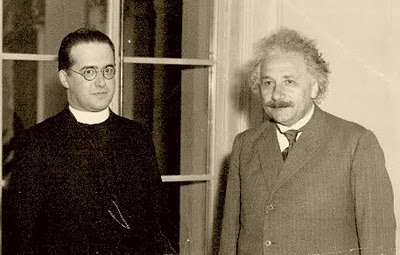}
\end{center}
\caption{Albert Einstein (1879-1955), ici avec Georges Lemaitre (1894-1966), était sceptique à propos de l'efficacité du libre marché pour des raisons indépendantes des externalités environementales.}
\label{fig:einstein}
\end{figure}

This crippling of individuals I consider the worst evil of capitalism. Our whole educational system suffers from this evil. An exaggerated competitive attitude is inculcated into the student, who is trained to worship acquisitive success as a preparation for his future career. 
\end{quote}

Albert Einstein proposait ensuite de considérer sérieusement une autre organisation de la société: 

\begin{quote}
I am convinced there is only one way to eliminate these grave evils, namely through the establishment of a socialist economy, accompanied by an educational system which would be oriented toward social goals. In such an economy, the means of production are owned by society itself and are utilized in a planned fashion. A planned economy, which adjusts production to the needs of the community, would distribute the work to be done among all those able to work and would guarantee a livelihood to every man, woman, and child. The education of the individual, in addition to promoting his own innate abilities, would attempt to develop in him a sense of responsibility for his fellow men in place of the glorification of power and success in our present society. \\
\end{quote}

\section{Remerciements}

Je remercie d'abord les coauteurs de \emph{Climat: 15 vérités qui dérangent}, de donner l'occasion de vulgariser l'état de la science du climat et de ses implications en économie. Je remercie plus particulièrement le professeur Markó de m'avoir présenté ses vues sur la climatologie au cours d'un séminaire qui s'est tenu le 10 avril 2012 à Schaerbeek. Je remercie \'egalement les personnes suivantes, sceptiques ou non sur le changement climatique anthropique ou sur l'efficacité du libre marché, pour des discussions utiles, des encouragements ou des commentaires sur des versions préliminaires de cet article: Pascal Acot, Normand Baillargeon, Sylvain Bouillon, Benjamin Bourgine, André Brahic, Fran\c{c}ois-Marie Br\'eon, Jean Bricmont, Fran\c{c}ois Brouyaux, Fabien Darrouzet, Anne De Rudder, Pierre Defrance, Emmanuel Dekemper, Nicolas Gauvrit, Georges Geuskens, S\'ebastien Henrotin, Jean-Lionel Lacour, Anabel-Lise Le Roux, Vincent Mathieu, Dominique Meeus, Frank Pattyn, Jean Pestieau, Romain Maggiolo, Gaia Pinardi, Lizz Printz, Xavier Urbain, François Vander Stappen, Filip Vanhellemont, Corinne Vigouroux, S\'ebastien Viscardy, Cedric Villani, Valérie Wilquet. Je remercie enfin et surtout Cristina Brailescu pour sa patience et son soutien.


\bibliography{markivbib}

\begin{thebibliography}{10}

\bibitem{marko_2013}
I.~Mark\'o, A.~Debeil, L.~Delory, S.~Furfari, D.~Godefridi, H.~Masson,
  L.~Myren, and A.~Pr\'eat, {\em Climat: 15 v\'erit\'es qui d\'erangent}.
\newblock Bruxelles: Texquis, 2013.

\bibitem{oreskes_2012}
N.~Oreskes, E.~M. Conway, and J.~Treiner, {\em Les marchands de doute}.
\newblock Paris: Le Pommier, 2012.

\bibitem{recherche_2013}
H.~Guillemot and S.~Aykut, ``Trois d\'ebats sur le climat,'' {\em La
  Recherche}, vol.~478, p.~73, 2013.

\bibitem{kaufman1992}
R.~Kaufman, ``Rio document spurs debate: Is science an ecological foe?,'' {\em
  The Scientist}, vol.~6, no.~15, 1992.

\bibitem{allegre2010}
C.~Allegre, {\em L'imposture climatique}.
\newblock Plon, 2010.

\bibitem{huet2010}
S.~Huet, {\em L'imposteur, c'est lui}.
\newblock Stock, 2010.

\bibitem{arrhenius_1897}
S.~Arrhenius, ``On the influence of carbonic acid in the air upon the
  temperature of the earth,'' {\em Publ. Astron. Soc. Pac.}, vol.~9, p.~14,
  1897.

\bibitem{arrhenius_1908}
S.~Arrhenius, {\em Worlds in the making: the evolution of the universe}.
\newblock Harper, 1908.

\bibitem{tyndall_1827}
J.~Tyndall, {\em Contributions to molecular physics in the domain of radiant
  heat}, vol.~7, pp.~421--424.
\newblock Longmans, Green, and co., 1872.

\bibitem{fourier_1827}
J.~Fourier, ``M\'emoire sur les temp\'eratures du globe terrestre et des
  espaces plan\'etaires,'' {\em M\'emoires de l'Acad\'emie Royale des Sciences
  de l'Institut de France}, vol.~7, pp.~569--604, 1827.

\bibitem{jacob_1999}
D.~Jacob, {\em Introduction to Atmospheric Chemistry}.
\newblock Princeton University Press, 1999.

\bibitem{IPCC2007}
IPCC, {\em Contribution of Working Group I to the Fourth Assessment Report of
  the Intergovernmental Panel on Climate Change}.
\newblock Cambridge University Press, 2007.

\bibitem{mann_1999}
M.~E. Mann, R.~S. Bradley, and M.~K. Hughes, ``Northern hemisphere temperatures
  during the past millennium: Inferences, uncertainties, and limitations,''
  {\em Geophys. Res. Lett.}, vol.~26, no.~6, 1999.

\bibitem{IPCC2001}
IPCC, {\em Climate Change 2001: The Scientific Basis, Contribution of Working
  Group I to the Third Assessment Report of the Intergovernmental Panel on
  Climate Change}.
\newblock Cambridge University Press, 2001.

\bibitem{mcintyre_2005}
S.~{McIntyre} and R.~{McKitrick}, ``Hockey sticks, principal components, and
  spurious significance,'' {\em Geophys. Res. Lett.}, vol.~32, no.~3, 2005.

\bibitem{mann_2004}
M.~E. Mann, R.~S. Bradley, and M.~K. Hughes, ``Corrigendum: Global-scale
  temperature patterns and climate forcing over the past six centuries,'' {\em
  Nature}, vol.~430, no.~6995, pp.~105--105, 2004.

\bibitem{jacoby_1997}
G.~C. Jacoby and R.~D. {D'Arrigo}, ``Tree rings, carbon dioxide, and climatic
  change,'' {\em P. Natl. Acad. Sci. USA}, vol.~94, no.~16, pp.~8350--8353,
  1997.

\bibitem{caillon_2003}
N.~Caillon, J.~P. Severinghaus, J.~Jouzel, J.-M. Barnola, J.~Kang, and V.~Y.
  Lipenkov, ``Timing of atmospheric {CO2} and antarctic temperature changes
  across termination {III},'' {\em Science}, vol.~299, no.~5613,
  pp.~1728--1731, 2003.

\bibitem{milankovic_1920}
M.~Milankovi\'{c} and {Jugoslavenska akademija znanosti i umjetnosti.}, {\em
  Th\'eorie math\'ematique des ph\'enom\`enes thermiques produits par la
  radiation solaire}.
\newblock Paris: Gauthier-Villars et Cie, 1920.

\bibitem{lorius_1990}
C.~Lorius, J.~Jouzel, D.~Raynaud, J.~Hansen, and H.~L. Treut, ``The ice-core
  record: climate sensitivity and future greenhouse warming,'' {\em Nature},
  vol.~347, no.~6289, 1990.

\bibitem{heimann_2008}
M.~Heimann and M.~Reichstein, ``Terrestrial ecosystem carbon dynamics and
  climate feedbacks,'' {\em Nature}, vol.~451, no.~7176, pp.~289--292, 2008.

\bibitem{global_2009}
U.~S. G. C.~R. Program, {\em Global Climate Change Impacts in the United
  States}.
\newblock Cambridge University Press, 2009.

\bibitem{Kerrick1998}
D.~M. Kerrick and K.~Caldeira, ``Metamorphic \{CO2\} degassing from orogenic
  belts,'' {\em Chemical Geology}, vol.~145, no.~3, 1998.

\bibitem{kerrick_2001}
D.~M. Kerrick, ``Present and past nonanthropogenic {CO2} degassing from the
  solid earth,'' {\em Reviews of Geophysics}, vol.~39, no.~4, 2001.

\bibitem{beck_2007}
E.-G. Beck, ``180 years of atmospheric
  {CO{\textless}SUB{\textgreater}2{\textless}/SUB{\textgreater}} gas analysis
  by chemical methods,'' {\em Energy \& Environment}, vol.~18, no.~2,
  pp.~259--282, 2007.

\bibitem{widory_2003}
D.~Widory and M.~Javoy, ``The carbon isotope composition of atmospheric {CO2}
  in paris,'' {\em Earth and Planetary Science Letters}, vol.~215, no.~1, 2003.

\bibitem{andrews_2010}
D.~G. Andrews, {\em An introduction to atmospheric physics}.
\newblock Cambridge; New York: Cambridge University Press, 2010.

\bibitem{rohde_2013}
R.~Rohde, R.~A. Muller, R.~Jacobsen, E.~Muller, and C.~Wickham, ``A new
  estimate of the average earth surface land temperature spanning 1753 to
  2011,'' {\em Geoinformatics \& Geostatistics: An Overview}, vol.~01, no.~01,
  2013.

\bibitem{hogg_2008}
A.~M. Hogg, ``Glacial cycles and carbon dioxide: A conceptual model,'' {\em
  Geophys. Res. Lett.}, vol.~35, no.~1, 2008.

\bibitem{IPCC2013}
IPCC, {\em Climate Change 2013: The Scientific Basis, Working Group I
  Contribution to the Fifth Assessment Report of the Intergovernmental Panel on
  Climate Change}.
\newblock Cambridge University Press, 2013.

\bibitem{liou_2002}
K.-N. Liou, {\em An Introduction to Atmospheric Radiation}.
\newblock Academic Press, 2002.

\bibitem{manabe_75}
S.~Manabe and R.~T. Wetherald, ``The effects of doubling the {CO}
  $_{\textrm{2}}$ concentration on the climate of a general circulation
  model,'' {\em Journal of the Atmospheric Sciences}, vol.~32, no.~1,
  pp.~3--15, 1975.

\bibitem{IPCC1997}
IPCC, {\em The Regional Impacts of Climate Change: An Assessment of
  Vulnerability}.
\newblock Cambridge University Press, 1997.

\bibitem{muller_2013}
R.~A. Muller, J.~Curry, D.~Groom, R.~Jacobsen, S.~Perlmutter, R.~Rohde,
  A.~Rosenfeld, C.~Wickham, and J.~Wurtele, ``Decadal variations in the global
  atmospheric land temperatures,'' {\em J. Geophys. Res-Atmos.}, vol.~118,
  no.~11, pp.~5280--5286, 2013.

\bibitem{muller_2012}
R.~A. Muller, ``The conversion of a climate-change skeptic,'' {\em The New York
  Times}, July 2012.

\bibitem{galilei_1957}
G.~Galilei, {\em Discoveries and opinions of Galileo.}
\newblock Garden City, {N.Y.}: Doubleday, 1957.

\bibitem{russell_1997}
B.~Russell and M.~Ruse, {\em Religion and science}.
\newblock New York: Oxford University Press, 1997.

\bibitem{Heymann-2010}
M.~Heymann, ``The evolution of climate ideas and knowledge,'' {\em Climate
  Change}, vol.~1, no.~4, pp.~581--597, 2010.

\bibitem{vonNeumann-1955}
J.~{von Neumann}, ``Can we survive technology?,'' {\em Fortune}, vol.~91,
  no.~6, 1955.

\bibitem{feynman_1974}
R.~Feynman, ``Cargo cult science,'' {\em Engineering and Science}, vol.~37,
  no.~7, 1974.

\bibitem{myren_2009}
L.~Myren and A.~Debeil, ``Have humans changed the climate? facts and
  consequences,'' {\em The Heartland Institute}, 2009.

\bibitem{nature_2011}
Editorial, ``Heart of the matter,'' {\em Nature}, vol.~475, no.~7357, 2011.

\bibitem{lewandowsky_2013}
S.~Lewandowsky, K.~Oberauer, and G.~E. Gignac, ``{NASA} faked the moon landing
  therefore, (climate) science is a hoax an anatomy of the motivated rejection
  of science,'' {\em Psychological Science}, vol.~24, no.~5, 2013.

\bibitem{hickman_2012}
L.~Hickman, ``Heartland institute compares belief in global warming to mass
  murder,'' {\em The Guardian}, 2012.

\bibitem{stern_2007}
A.~Benjamin, ``Stern: Climate change a 'market failure','' {\em The Guardian},
  2007.

\bibitem{Block_2006}
W.~Block, ``Defending the speculator,'' {\em Mises Institute}, 2010.

\bibitem{smith_1904}
A.~Smith, {\em An Inquiry into the Nature and Causes of the Wealth of Nations}.
\newblock Edwin Cannan ed., 1904.

\bibitem{meeropol_2004}
M.~Meeropol, ``Another distortion of adam smith: The case of the {"Invisible}
  hand",'' {\em {PERI} Working Papers}, 2004.

\bibitem{vergara_2002}
F.~Vergara, {\em Les fondements philosophiques du lib\'eralisme}.
\newblock La D\'ecouverte, 2002.

\bibitem{stiglitz_2006}
D.~Altman, ``Q and a with joseph stiglitz in managing globalization,'' {\em The
  International Herald Tribune}, 2006.

\bibitem{pigou_1920}
A.~C. Pigou, {\em The Economics of Welfare}.
\newblock Palgrave Macmillan, 2013.

\bibitem{hayek_2001}
F.~A.~v. Hayek, {\em The road to serfdom}.
\newblock London: Routledge, 2001.

\bibitem{coase_1960}
R.~H. Coase, ``The problem of social cost,'' {\em Journal of Law \& Economics},
  vol.~3, p.~1, 1960.

\bibitem{stigler_1987}
G.~Stigler, {\em The theory of price}.
\newblock Macmillan, 1987.

\bibitem{Barnett_2009}
A.~H. Barnett and B.~Yandle, ``The end of the externality revolution,'' {\em
  Social Philosophy and Policy}, vol.~26, pp.~130--150, 7 2009.

\bibitem{brohe_2008}
A.~Broh\'e, {\em Les march\'es de quotas de {CO2}}.
\newblock Bruxelles: Larcier, 2008.

\bibitem{bernier_2008}
A.~Bernier, {\em Le climat, otage de la finance}.
\newblock [Paris]: Mille et une nuits, 2008.

\bibitem{quenault_2009}
B.~Quenault, ``Lectures: Aur\'elien bernier, 2008, le climat otage de la
  finance, essai, mille et une nuits, paris, 164 p.,'' {\em D\'eveloppement
  durable et territoires}, 2009.

\bibitem{lindzen_2013}
R.~Lindzen, ``Science in the public square: Global climate alarmism and
  historical precedents,'' {\em Journal of American Physicians and Surgeons},
  vol.~18, no.~3, 2013.

\bibitem{halimi_2004}
S.~Halimi, {\em Le grand bond en arriere}.
\newblock Paris: Fayard, 2004.

\bibitem{Einstein_1949}
A.~Einstein, ``Why socialism?,'' {\em Monthly Review}, 1949.

\end{thebibliography}

\end{document}